\documentclass[twocolumn,english,amssymb,prl,superscriptaddress,notitlepage]{revtex4-2}
\usepackage{amsmath,amssymb,graphicx}
\usepackage{multirow}
\usepackage{fancyhdr,bm}
\usepackage{booktabs} 
\usepackage{braket}
\usepackage{titlesec}
\titleformat{\paragraph}
{\normalfont\normalsize\bfseries}{\theparagraph}{1em}{}
\titlespacing*{\paragraph}
{0pt}{3.25ex plus 1ex minus .2ex}{1.5ex plus .2ex}
 
\usepackage{color}
\usepackage[colorlinks=true, pdfstartview=FitV, linkcolor=blue, citecolor=blue, urlcolor=blue]{hyperref}
\nonstopmode

\begin{document}
%\title{Topological Pair Density Wave in Rhombohedral Graphene}
\title{Quarter Metal Superconductivity}
\author{Chiho Yoon}
\affiliation{Department of Physics, University of Texas at Dallas, Richardson, Texas 75080, USA}
\author{Tianyi Xu}
\affiliation{Department of Physics, University of Texas at Dallas, Richardson, Texas 75080, USA}
\author{Yafis Barlas}
\affiliation{Department of Physics, University of Nevada, Reno, Nevada 89557, USA}
\author{Fan Zhang}
\affiliation{Department of Physics, University of Texas at Dallas, Richardson, Texas 75080, USA}
\begin{abstract}
We investigate the recently discovered multiple superconducting states in rhombohedral graphene quarter metal.
We demonstrate that one of these states features a single-spin, single-valley, single-band, single-Fermi-pocket parent state 
and is most likely a chiral topological pair-density wave, 
marked by a threefold symmetry that may not be spontaneously broken, 
unpaired Majorana zero modes at edges, vortices, and dislocations, and an anomalous intrinsic superconducting diode effect.
\end{abstract}
\date{\today} 
\maketitle

\indent\textcolor{blue}{\em Introduction.}---%
Two-dimensional semiconductors have emerged as a central focus in condensed matter physics, 
driving discoveries in fundamental principles, quantum matter, and functional devices. 
A prime example is ABC or rhombohedral stacked multilayer graphene, 
which has been theoretically predicted~\cite{Zhang2010a,Zhang2011a} and experimentally confirmed~\cite{Weitz2010a, Martin2010a,
Lui2011a, Bao2011a, Zhang2011b, Zou2013a,
Velasco2012a, Freitag2012a, Bao2012a, Shi2020a, Han2024c, Liu2024b, 
Han2024a, Sha2024a, Winterer2024a, Geisenhof2021a, 
Zhou2021a, Barrera2022a, Seiler2022a, Zhou2022a, Yang2024a,
Zhou2021b, Zhang2023a}
as ``{fertile ground for new physics}''~\cite{Zhang2010a}, 
establishing a new paradigm for the interplay of symmetry, topology, and interaction. 
This family of naturally occurring crystals exhibits electrically tunable band gaps~\cite{Zhang2010a, Zou2013a, Shi2020a, Koshino2009a}, 
high material quality, device reconfigurability, and exceptionally rich many-body physics.

At charge neutrality, the flat touching bands and large winding numbers of rhombohedral graphene 
give rise to spontaneous chiral symmetry breaking and valley-projected topological insulators. 
%% Zhang2012b, Zhang2015a
These include layer antiferromagnetic ground states 
and ferroelectric quantum anomalous Hall (QAH) states~\cite{Zhang2011a, Jung2011a, Zhang2012a}   
(with Chern number scaling with layer thickness), which have been verified by subsequent experiments~\cite{Velasco2012a, Freitag2012a, Bao2012a, Shi2020a, Han2024c, Liu2024b, Han2024a, Sha2024a, Winterer2024a, Geisenhof2021a}. 
Furthermore, the fine structure of band touching exhibits van Hove singularities (vHS)~\cite{Zhang2010a, Koshino2009a}, 
which can be continuously tuned to higher densities by an electric field. 
The vHS drive isospin Stoner magnetism~\cite{Zhou2021a, Barrera2022a, Seiler2022a, Zhou2022a, Yang2024a}, 
novel superconductivity~\cite{Zhou2022a, Yang2024a, Zhou2021b, Zhang2023a}, 
and possibly anomalous Wigner-Hall crystallization~\cite{Seiler2022a, Dong2024c}.
%% Soejima2024a
Additionally, moiré engineering has enabled the realization of flat Chern bands and the fractional QAH effect~\cite{Chittari2019a, Lu2024a}.
%% Chen2019a

Recently, a tetralayer experiment~\cite{Han2024b} discovered two fluctuating superconducting states 
that can be stabilized by strong magnetic fields in the spin- and valley-polarized Stoner phase, referred to as quarter metal (QM). 
This counterintuitive finding has inspired several theories~\cite{Qin2024a,Wang2024a,Yang2024b,Geier2024a,Chou2024a,Kim2025a,Jahin2024a} 
proposing pairing mechanisms for chiral superconductivity. 
Here, we present a theory for QM superconductivity that does not rely on such mechanisms. 
We first construct a QM model informed by first-principles calculations, 
revealing the density of states (DOS) and Fermi surface geometry, 
which closely resemble the experimental data. 
Next, we classify the possible pairing channels and demonstrate that 
the most natural one is the Fulde-Ferrell pair density wave (PDW)~\cite{Fulde1964a}, 
which exhibits reduced translational symmetry but unbroken rotational symmetry. 
Our superfluidity stiffness calculations further suggest that the quantum geometric contribution is negligible, 
validating a chiral PDW arising from only one band and one Fermi pocket. 
We predict two key characteristics of this simplest topological PDW: 
Majorana zero modes bound not only to edges and vortices but also to dislocations of the PDW order parameter, 
and an anomalous intrinsic superconducting diode effect.

\begin{figure}[t!]
\centering
\includegraphics[width=1.0\linewidth]{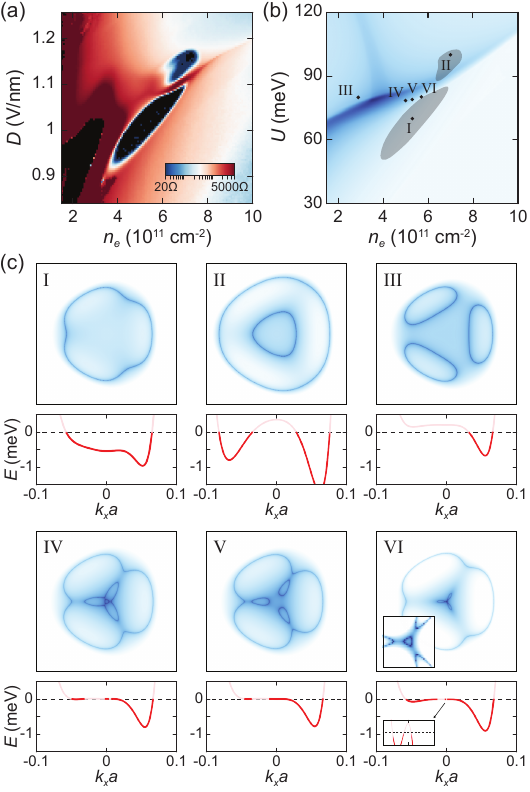}
\caption{(a) $R_{xx}$ data of tetralayer rhombohedral graphene at $B_z=0.1$~T adapted from a recent experiment~\cite{Han2024b}.
(b) DOS map based on first-principles calculations for tetralayer rhombohedral graphene as a function of 
the electron density $n_{e}$ and the screened potential difference between the top and bottom layers $U$. 
The dark lines correspond to vHS and their vicinities. 
The grey regions labeled by I and II in (b) correspond to the two superconducting states in (a).
(c) Fermi surfaces (top) and band structures (bottom) for various states labeled in (b). The Fermi energies are all shifted to zero.}
\label{fig1}
\end{figure}

\indent\textcolor{blue}{\em Parent state modeling.}---%
We first use {\it ab initio} density functional theory to calculate the electronic band structure of pristine tetralayer rhombohedral graphene. 
We then fit the low-energy bands to an eight-band tight-binding model (with 2 sublattices for each of the 4 layers) 
to extract the hopping parameters~\cite{Zhang2010a}. These two steps (see End Matter) are necessary 
because the trigonal warping and fermiology at low electron densities are sensitive to the values of the hopping parameters,
which are thickness dependent. 
To compare with the experimental $R_{xx}$ data~\cite{Han2024b} reproduced in Fig.~\ref{fig1}(a), 
we use the eight-band model to calculate the DOS of isospin ($K$, $\uparrow$) as a function of the electron density $n_{e}$ 
and the potential difference between the top and bottom layers $U$, as plotted in Fig.~\ref{fig1}(b). 
Fig.~\ref{fig1}(c) further plots the Fermi surfaces and band structures for six selected ($n_{e}$, $U$) pairs labeled in Fig.~\ref{fig1}(b).
In region I, the Fermi surface is a trigonally warped electron pocket. 
In region II, the Fermi surface becomes annular. 
In between, the Fermi surface is complicated by the multiple appearances of saddle points 
associated with the evolution of the inner hole pocket. 
The corresponding vHS and their vicinities form a dark line in Fig.~\ref{fig1}(b). 
All these features match well with the $R_{xx}$ and fermiology data in experiment~\cite{Han2024b}.

Similar to rhombohedral graphene of different thickness, a larger $U$ can push the vHS into higher $n_{e}$.
Along this vHS line, the inner Fermi-surface complex is strongly warped, and the band edge is partially ultra-flat, 
as shown in Fig.~\ref{fig1}(c) (IV–VI).   
This implies that pairing between states with opposite intra-valley momenta is disfavored, 
possibly explaining the experimentally observed disruption of superconductivity between the I and II regions. 
In region III of relatively larger $U$ and lower $n_{e}$, the Fermi surface has three isolated pockets around the $K$ point. 
Similarly to the states along the vHS line, pairing within the $K$ valley is almost impossible, 
as revealed by the band structure in Fig.~\ref{fig1}(c) (III). Indeed, no superconductivity has been found experimentally in this region~\cite{Han2024b}. 
However, inter-valley pairing is possible for the half-metal counterpart of this three-pocket state. 
In experiment~\cite{Han2024b}, a superconducting state emerges at nearly twice the density of this state, 
and it disappears beyond a small magnetic field that induces a valley splitting; 
electrons at the two valleys have large but opposite orbital magnetic moments~\cite{Zhang2011a}. 

The analysis above validates the use of single-isospin band states as the spin- and valley-polarized QM phase relevant to the experiment~\cite{Han2024b}.
Further support exists in the Hartree-Fock theory of such Stoner-like phases. 
The interactions primarily split the isospin bands in energy rather than reshaping their dispersions. In addition, 
the energy splitting does not have a dramatic density dependence within the QM phase. 
Thus, we can use an isospin-dependent mean-field $H_{\rm int}$ acting on the low-energy sublattices 
1A and 4B ($\sigma_z\!=\!\pm 1$) to model the QM phase:
\begin{eqnarray}
\label{eq:effective_qm_int}
H_{\rm int} = V_{\rm int} \sigma_{z} ( 1- \tau_{+} s_{+}),
\end{eqnarray}
where $\tau_{+} \!=\! (\tau_{0} + \tau_{z})/2$ with $\tau_z \!=\!\pm 1$ the $K$/$K'$ valleys 
and $s_{+} \!=\! (s_{0} + s_{z})/2$ with $\bm s$ the spin Pauli matrices.
We can estimate the constant $V_{\rm int}$ by identifying the electron density required 
to enter the half-metal phase at a fixed gate-displacement field. 
The quantum oscillation data~\cite{Han2024b} shows that the half-metal phase starts to emerge
at $n_e\!=\!1.35 \!\times\! 10^{12}$~cm$^{-2}$ at $D\!=\!1$~V/nm
(corresponding to $U\!=\!65$~meV in Fig.~\ref{fig1}), indicating $V_{\rm int}\!=\!24$~meV.

\begin{table}[b!]
\renewcommand{\arraystretch}{1.75} 
\setlength{\tabcolsep}{4pt} 
\centering
\caption{Character table for the irreps associated with the $K$-valley $\uparrow$-spin PDW in the $C_{3z}''$ extended point group, 
along with the corresponding order parameters up to first order in intra-valley momentum.
Here, $\omega = e^{2\pi i/3}$ and $k_{\pm}= k_x \pm i k_y$.}
\vspace{0.05in}
\begin{tabular}{c|c|c|c|c}
\hline\hline
Irrep		&	$C_{3z}^{\rm{A}}$ & $C_{3z}^{\rm{B}}$  & $C_{3z}^{\rm{C}}$    & Order Parameter \\
\hline
$A_{{K}}$ 		& 1				& $\omega$		& $\bar{\omega}$	& $\Delta_{A}^{(0)} i\sigma_{y}+\Delta_{A}^{(-)} k_{-} \sigma_{+} +\Delta_{A}^{(+)}k_{+} \sigma_{-}$ \\
$E_{{K}}$ 		& $\omega$		& $\bar{\omega}$	& 1				& $\Delta_{E}^{(+)} k_{+} \sigma_{+} + \Delta_{E}^{(-)} k_{-}\sigma_{x}$ \\
$\bar{E}_{{K}}$	& $\bar{\omega}$	& 1				& $\omega$		& $\Delta_{\bar{E}}^{(-)} k_{-} \sigma_{-} + \Delta_{\bar{E}}^{(+)} k_{+}\sigma_{x}$ \\
\hline\hline
\end{tabular}
\label{table}
\end{table}

\begin{figure}[t!]
\centering
\includegraphics[width=1\linewidth]{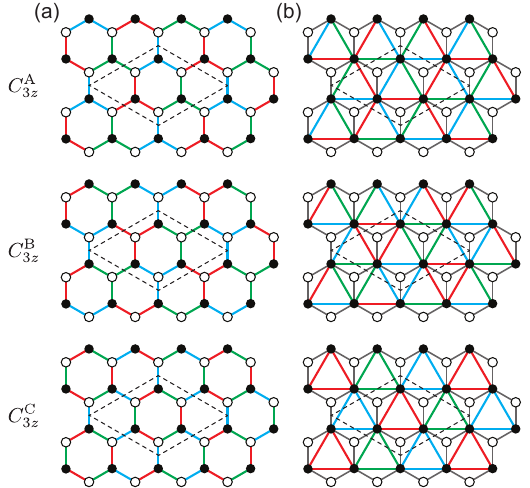}
\caption{Schematics of the three nearest (a) 1A-4B and (b) 1A-1A pairing channels for the $K$ valley PDW states,  
which respect the $C_{3z}^{\rm{A}}$, $C_{3z}^{\rm{B}}$, and $C_{3z}^{\rm{C}}$ symmetries, respectively.
The solid and open circles denote the low-energy sublattices 1A (top layer) and 4B (bottom layer), respectively.
The red, green, and blue lines sketch the pairing with the phase ($2\bm{K}\cdot \bm{r}$) being $0$, $2\pi/3$, and $4\pi/3$, respectively.
The dashed lines show the $(\sqrt{3}\times\sqrt{3})R30^{\circ}$ supercells of the PDW states. 
These PDW states belong to the wallpaper group $P_3$.}
\label{fig2}
\end{figure}

\indent\textcolor{blue}{\em PDW classification.}---%
We focus on possible superconducting pairing channels within the ($K$, $\uparrow$) isospin band 
and ignore those involving any other isospin band, given that 
%%only one electron band is populated in the parent-state QM, and that 
other bands are at least $V_{\rm{int}}\!=\!24$~meV away.
%%which is significantly larger than the superconducting order parameter $<1$~meV.
This suggests PDW states, as the Cooper pairs forming from the tiny Fermi surface within the $K$ valley   
must carry a giant momentum $2{K}\!=\!({8\pi}/{3a},0)$, where $a$ is the graphene lattice constant. 
The most natural state is thus the simplest Fulde-Ferrell-type~\cite{Fulde1964a} commensurate PDW, 
which has a plane-wave-like superconducting {\it phase} modulation $e^{2i\bm{K}\cdot \bm{r}}$ or $e^{i\bm{K}'\cdot \bm{r}}$~\footnote{$2{\bm K}$ and ${\bm K}'$ differ by a reciprocal lattice vector.}. 
To illustrates this spatially varying phase, Fig.~\ref{fig2} sketches six examples:
translation by a graphene primitive lattice vector $\bm{a}_{1}$ or $\bm{a}_{2}$ results in
a phase shift of the order parameter by $2\pi/3$ or $4\pi/3$, respectively, 
while leaving the amplitude unchanged. As a consequence of the PDW order, the unit cell is extended by three times.
In general, the anisotropy of the trigonally-warped Fermi surface 
allows for incommensurate PDW states with the pair momenta deviating from $2K$.
However, the QM Fermi surface is tiny; the Fermi wavevector is $\sim {2\pi}/{100a}$ in Fig.~\ref{fig1}(c),  
indicating that the deviation may only be observed over a length scale much larger than 100 unit cells and thus negligible. 
 
We now classify the PDW order parameters based on the symmetries of the parent-state QM. 
The applied gate-displacement field explicitly breaks inversion symmetry, 
and the QM spontaneously breaks time-reversal and mirror symmetries. 
(All these symmetries preserve each sublattice but relate the two valleys.)
Remaining are two lattice translational symmetries and one threefold rotational symmetries, $T_{1,2}$ and $C_{3z}$.
Different from the classification of ordinary superconducting states, 
here the translational symmetries need to be taken into account, in addition to the point group symmetry,
as the PDW states spontaneously break not only the gauge symmetry but also $T_{1,2}$.
Thus, we classify the irreducible representations (irreps)~\cite{Tinkham1964a} of a so-called \textit{extended} point group~\cite{Venderbos2016a} $C_{3z}''$, 
generated by $C_{3z}$ and $T_{1,2}$. As illustrated in Fig.~\ref{fig2}, $T_{1}^{2} = T_{2}$ and $T_{1}^{3} = I$ hold for the extended unit cells.
As a result, the extended point group $C_{3z}''$ is isomorphic to the abelian group $\mathbb{Z}_{3} \times \mathbb{Z}_{3}$, 
which has nine irreps in total.
Among them, three do not break $T_{1,2}$ (but may break $C_{3z}$) 
and correspond to valley-singlet pairing between the two valleys,
three characterize the valley-triplet spin-triplet PDW within the $K$ valley, as presented in Table~\ref{table}, 
and the remaining three characterize the counterpart PDW within the $K'$ valley. 

The three irreps are distinguished by the $C_{3z}$, $T_{1}C_{3z}$, and $T_{2}C_{3z}$ symmetry operations.
The $C_{3z}$ rotation axis of rhombohedral graphene can be placed across the A, B, or C (hollow) sublattice.  
In fact, assuming $C_{3z} \equiv C_{3z}^{\rm A}$ is the rotational operation around A, 
$T_{1}C_{3z} \equiv C_{3z}^{\rm B}$ and $T_{2}C_{3z}  \equiv C_{3z}^{\rm C}$ are the same rotational operation as $C_{3z}$ but around B and C, respectively.
In the presence of the $T_{1,2}$ symmetries, the three rotational operations are equivalent.
In the PDW states breaking the $T_{1,2}$ symmetries, however, they become inequivalent. 
Remarkably, the allowed PDW states are special orders that do not break the $C_{3z}$ symmetry (as $\bm{K}$ is $C_{3z}$ invariant) 
at the expense of breaking the $T_{1,2}$ symmetries. As shown in Table~\ref{table}, 
each irrep must respect one of the $C_{3z}^{\rm A}$, $C_{3z}^{\rm B}$, and $C_{3z}^{\rm C}$ symmetries~\footnote{where  $C_{3z}^{\rm A}$ is represented as $e^{-i\frac{\pi}{3}s_z}(\sigma_+ + e^{i\frac{2\pi}{3}\tau_z}\sigma_-)$ in the low-energy sublattice basis formed by 1A and 4B.}. 
This indicates that nematic superconductivity is impossible (suppressed) at the mean-field level for commensurate (incommensurate) PDW.

\indent\textcolor{blue}{\em One-band character.}---%
Based on the classification of the irreps, the corresponding order parameters up to first order in $\bm k$, 
the momentum deviation from the $K$ point, are listed in Table~\ref{table}. 
The Taylor expansion is valid given the small size of the Fermi surface. 
The pairing channels within the 1A (4B) sublattice are proportional to $\sigma_{+}$ ($\sigma_{-}$), 
and they exhibit $k_x\pm i k_y$ chiral superconductivity due to the triplet nature and the $C_{3z}$ symmetry.
By contrast, inter-sublattice pairing channels are proportional to $\sigma_{x,y}$.
Although quantitative analysis depends on a specific pairing mechanism, the pairing channels within the 1A sublattice, {\it i.e.}, 
$\Delta_{A}^{(-)} k_{-} \sigma_{+}$ and $\Delta_{E}^{(+)} k_{+} \sigma_{+}$, are anticipated to dominate, 
because the low-energy states at the conduction-band edge are polarized to the 1A sublattice by the strong gate-displacement field.

It is still essential to validate that the QM superconductivity can be accurately modeled using only the lowest conduction band or the 1A 
sublattice~\footnote{As in the wallpaper group $P_3$, the symmetry should be $C_{3z}$ instead of $C_{6z}$, although the 1A sublattice is triangular.}, 
given the large valley Chern number of each isospin band~\cite{Zhang2011a} 
and the flat band edge near the Fermi level~(Fig.~\ref{fig1}). 
Quantum geometry, as a multi-band effect, may significantly contribute to the superfluid stiffness $\rho_s$~\cite{Peotta2015a,Rossi2021a, Torma2022a}. 
For instance, for an isolated flat band with vanishing intra-band contribution, $\rho_s$ is bounded below by its Chern number~\cite{Peotta2015a}. 
Here we directly calculate $\rho_s$ using a linear response theory 
for the superconducting state I with $\Delta_{A}^{(-)}\!=\!10$~meV (see End Matter). 
Our calculations obtain $\rho_s\!=\!1.05$~meV when using the eight-band model
and $\rho_s\!=\!1.08$~meV when using the lowest conduction band dispersion. 
This implies that the quantum geometric contribution to $\rho_s$ is negligible. 
Indeed, the strong displacement field substantially spreads the Berry curvature 
to the momentum space~\cite{Zhang2011a} and reduces the conduction-band Berry flux below the Fermi level.
Therefore, the QM superconductivity can be studied accurately using the single-band model. 
Furthermore, we find that trigonal warping and order parameter winding destabilize the chiral $p$-wave PDW, 
contributing $-5.2\%$ and $-7.7\%$ to $\rho_s$, respectively.

\begin{figure}[t!]
\centering
\includegraphics[width=1.0\linewidth]{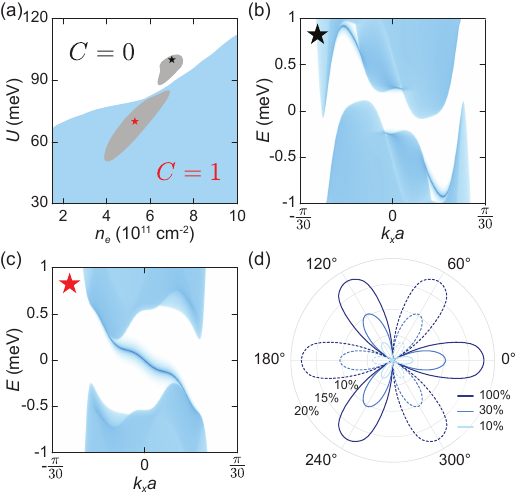}
\caption{(a) Topological phase diagram for the application of $\Delta_{A}^{(-)} = 10$~meV to the QM states in Fig.~\ref{fig1}(b). 
(b, c) Zigzag edge-state spectra for the two states marked in (a).
(d) Calculated zero-temperature diodicity for the superconducting state I in Fig.~\ref{fig1}(b) 
at $10\%$, $30\%$, and $100\%$ trigonal warping strengths.
$\theta=0$ ($\pi$) refers to the KM (K$\Gamma$) direction.}
\label{fig3}
\end{figure}

\indent\textcolor{blue}{\em Topological PDW.}---%
Chiral BCS superconductivity is characterized by a nontrivial Chern number. 
To explicitly demonstrate the topological nature of our chiral $p$-wave PDW, 
we lattice-regularize $H_{\rm int}$ in Eq.~(\ref{eq:effective_qm_int}) along with the order parameter $\Delta_{A}^{(-)} k_{-} \sigma_{+}$,  
and construct the BdG Hamiltonian for the eight-band tight-binding model (see End Matter). 
For simplicity, we choose $\Delta_{A}^{(-)}$ to be $10$~meV 
and assume such QM superconductivity for the entire region of Fig.~\ref{fig1}(b).  
As depicted in Fig.~\ref{fig3}(a), the single-Fermi-pocket region I exhibits a Chern number $C=1$, 
while the annular-Fermi-surface region II is topologically trivial due to the opposing contributions from the electron and hole pockets. 
Markedly, region I features a chiral Majorana edge mode, while region II does not, 
as shown in Fig.~\ref{fig3}(b, c)~\footnote{Additional trivial zero modes may arise from the large valley Chern numbers: $3$ or $5$ for ($K$, $\uparrow$) and $\pm 4$ for other isospins.}. 

The topological transition occurs at the vHS line. Indeed, the order parameter vanishes at the $K$ point, 
and the QM states at the vHS line feature Fermi pockets near the $K$ point, as shown in Fig.~\ref{fig1}(c). 
This results in nodal BdG spectra, which may disfavor superconductivity, 
in contrast to the widely studied vHS enhancing superconductivity~\cite{Nandkishore2012a}. 
Thus, the experimentally observed disruption of superconductivity between regions I and II  
is likely evidence of our predicted chiral $p$-wave PDW. 

Next, we examine topological point defects in the superconducting region I, such as vortex, dislocation, and disclination, 
as the PDW spontaneously breaks or reduces the gauge, translational, and rotational symmetries.  
Consider the following minimal model~\cite{Alicea2012a}
\begin{equation}
H_{\rm d}\!=\!\!\int\!\!d^2 r \bigg[ \frac{\Delta}{2} e^{i ({\bm K}' \cdot \bm r -\theta)} \psi(\partial_r \!-\! \frac{i\partial_\theta}{r})\psi 
   - \mu \psi^\dagger \psi + {\rm h.c.}\bigg].\label{defect}
\end{equation}
The band energy term is ignored due to the slowly varying nature of the chemical potential $\mu (r)$, 
which switches sign at a circular interface $r=r_0$. 
The $r>r_0$ region is the chiral $p$-wave PDW with $\Delta=\Delta_{A}^{(-)}$ and $\mu (r)>0$, 
whereas the $r<r_0$ core is a vacuum state with $\mu (r)<0$. 
It is convenient to gauge the factor $e^{i ({\bm K}' \cdot \bm r -\theta)}$ away 
by defining $\psi=e^{i(\theta - {\bm K}'\cdot \bm r)/2}\psi'$. 
For modes localized at the defect, it suffices to replace $r\rightarrow r_0$. 
One then finds that the energies of the defect states read
\begin{align}
   E=\frac{n|\Delta|}{r_0}, 
\end{align}
with $n$ is a (half) integer if $\psi'$ exhibits (anti-) periodic boundary condition upon encircling the defect.
 
Notably, the $n\!=\!0$ state is an unpaired Majorana mode. 
Since $\psi$ must be single-valued, 
$(2\pi i)^{-1}\oint_{\gamma} \psi^{-1} d\psi$ must be an integer. 
For the ordinary chiral $p$-wave case, the PDW phase $\bm K'\!\cdot\!\bm r$ is absent in Eq.~(\ref{defect}),  
and $\psi'(\theta+2\pi)\!=\!-\psi'(\theta)$ given $\psi\!=\!e^{i\theta/2}\psi'$.
Threading a flux quantum ${hc}/{2e}$ through the defect core introduces a vortex in the pair field ($\Delta\!\rightarrow\!\Delta e^{i\theta}$),   
leading to $\psi'(\theta+2\pi)\!=\!\psi'(\theta)$ and thereby the emergence of the Majorana zero mode.

The PDW phase ${\bm K}'\!\cdot\!\bm r$ does not induce any phase winding around the defect unless a dislocation is present.
This topological excitation is characterized by the failure of closure in the displacement field 
${\bm u}({\bm r})$:~$\oint_{\gamma} d{\bm u}({\bm r}) \!=\! \bm b$, 
where $\bm b$ is the Burgers vector and $\bm b\!\cdot\!{\bm K}'$ is an integer multiple of $2\pi$~\cite{Chaikin1995a}, 
given that ${\bm K}'$ is a primitive reciprocal vector of the PDW.
This implies that a unit dislocation produces the same phase winding as the magnetic flux quantum through the vortex. 
As a result, the PDW dislocation binds a Majorana zero mode even in the absence of magnetic flux. 
In general, an unpaired Majorana emerges only when the sum of flux quanta and dislocation units at the defect is an odd integer. 
This new finding holds true for any chiral topological PDW.

By contrast, a unit dislocation of the graphene lattice does not produce a Majorana zero mode,
as the lattice Burgers vector is not a primitive vector of the PDW supercell.  
Unlike dislocations, disclinations are much more costly in energy~\cite{Chaikin1995a}. 
The Frank angle $\Omega=2\pi/3$ is expected to produce a $e^{i\Omega/2}$ twist in the boundary condition of $\psi'$  
and thus no Majorana zero mode~\cite{Teo2013a}. 
It will be interesting to explore the possible proliferation of Majoranas and dislocation-mediated melting of PDW.

\indent\textcolor{blue}{\em PDW diode effect.}---%
Finally, we show that the QM superconductivity exhibits an intrinsic superconducting diode effect (SDE) 
without the need of a Josephson junction~\cite{Ando2020a,Lin2022a,He2022a,Yuan2022a,Daido2022a},
because both time-reversal and inversion symmetries are broken. 
This non-reciprocity in superconducting transport can be characterized by the diodicity, 
$\eta(\theta)=[I_c(\theta)-I_c(\theta+\pi)] /{\rm max}\{I_c(\theta), I_c(\theta+\pi)\}$, 
where $I_c(\theta)$ and $I_c(\theta+\pi)$ are the forward and backward critical supercurrents, 
and $\theta=0$ ($\pi$) refers to the KM (K$\Gamma$) direction.  
Fig.~\ref{fig3}(d) shows the calculated zero-temperature diodicity for the superconducting state I 
at three different strengths of trigonal warping (see End Matter).
$\eta$ can reach $18\%$ in the KM and K$\Gamma$ directions but with opposite signs. 
The three-fold angular dependence of $\eta(\theta)$ reflects the unbroken $C_{3z}$ symmetry of PDW.
Moreover, $\eta(\theta)$ would vanish if valley polarization or trigonal warping was absent. 
This indicates that PDW or Fermi-surface anisotropy itself cannot render SDE.
Observing the predicted SDE in experiment can directly validate the valley-triplet nature of QM superconductivity. 
Noticeably, this {\it spin-polarized} intrinsic SDE is {\it anomalous} because neither magnetic field nor spin-orbit coupling is required, 
different from the previous findings~\cite{He2022a,Yuan2022a,Daido2022a}. 

\indent\textcolor{blue}{\em Discussion.}---%
When time-reversal or inversion symmetry is present, 
quasiparticle states of $\pm\epsilon_{\bm{k}}$ come in pair. 
Thus, the BdG spectrum develops a global gap once a direct gap opens (always at $\epsilon\!=\!0$).
By contrast, both symmetries are broken in the $K$-valley PDW, 
and the BdG spectrum satisfies $\epsilon_{\bm{K}+\bm{k}}\!=\!-\epsilon_{\bm{K}-\bm{k}}$ (Fig.~\ref{fig3}).
As a result, even though the order parameter induces a gap at each momentum,
the global BdG spectrum may remain gapless due to a negative indirect gap (NIG)~\cite{Halperin1968a}. 
In this scenario, a Bogoliubov Fermi surface may form, or more likely superconductivity is suppressed.
Figure~\ref{figEM2} presents the NIG maps of the BdG spectra with $\Delta_{A}^{(-)}\!=\!0$ and $10$~meV. 
Indeed, valley polarization and trigonal warping give rise to the dominance of NIG over $1$~meV. 
Evidently, the two maps share similar patterns, and separate superconducting regions do emerge near $5\!\times\! 10^{11}$~cm$^{-2}$.
These suggests that the revealed NIG and vHS disfavor chiral PDW or enhance its nodal nature.

In summary, we have shown that the QM in region I can be described by a single-isospin, single-band, single-pocket model. 
The superconducting state must be spin-triplet, valley-triplet, and odd-parity with respect to the momentum deviation from the $K$ point.
The `odd-parity' property reflects fermionic exchange statistics and the Fulde-Ferrell-type PDW. 
The small size of the Fermi surface favors $p$-wave pairing, 
while the unbroken $C_{3z}$ symmetry enforces its chiral nature. 
We anticipate that the PDW patterns, along with unpaired Majoranas at vortices and PDW dislocations, can be observed in STM experiments~\cite{Hamidian2016a,Liu2021a,Gu2023a,Deng2024a}. 
Additionally, the anomalous SDE is expected to be optimized as a function of charge density, electric field, and layer thickness.

\indent\textcolor{blue}{\em Acknowledgement.}---%
We are grateful to Long Ju for sharing with us their experimental data 
and for many fruitful discussions. We thank Huaixuan Li for his support with our first-principles calculations.
This work was supported by NSF under Grants No. DMR-1945351, No. DMR-2324033, and No. DMR-2414726. 
We acknowledge the Texas Advanced Computing Center for providing resources 
that have contributed to the research results reported in this work. 

\newpage
\bibliographystyle{apsrev4-2_edit_200521}
\bibliography{QMSC_ref}{}

\clearpage
\begin{appendix}

\begin{center}
\textbf{End Matter}
\end{center}

\begin{figure}[b]
\centering
\includegraphics[width=0.7\linewidth]{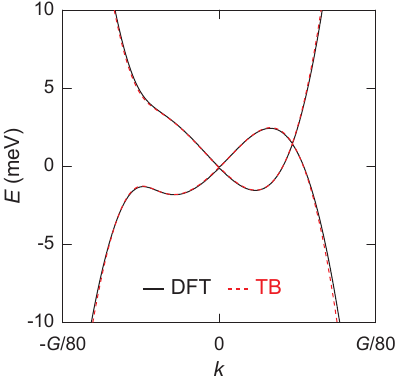}
\caption{Low-energy band structures of tetralayer rhombohedral graphene obtained from 
our bare DFT calculations (solid black) and our DFT-informed effective model (dashed red).}
\label{figEM1}
\end{figure}

\indent\textcolor{blue}{\em DFT and model.}---%
To obtain the low-energy electronic band structures of tetralayer rhombohedral graphene, 
we perform density functional theory (DFT) calculations by using the Vienna {\it ab initio} simulation package. 
The DFT-D3 functional is used to describe van der Waals interactions, 
and the interlayer distance is set to be the experimental value $3.35$~\AA~without relaxation.
We use the low-energy bands to fit the parameters of the phenomenological eight-band (per isospin) model  
using the sublattice basis \{1A, 1B, 2A, 2B, 3A, 3B, 4A, 4B\}~\cite{Zhang2010a}:
\begin{eqnarray}
&H = 
\begin{pmatrix}
H_{00}+\frac{U}{2}+\delta			& H_{01}			& H_{02}			& 0		\\
H_{01}^{\dagger}	& H_{00}+\frac{U}{6}			& H_{01}			& H_{02}	\\
H_{02}^{\dagger}	& H_{01}^{\dagger}	& H_{00}-\frac{U}{6}			& H_{01}	\\
0				& H_{02}^{\dagger}	& H_{01}^{\dagger}	& H_{00}-\frac{U}{2}+\delta	\\
\end{pmatrix}\nonumber \\
&H_{00} \!=\! 
\begin{pmatrix}
0		 		& \gamma_{0} f	 \\
\gamma_{0} f^{*}	& 0
\end{pmatrix}
H_{01} \!=\! 
\begin{pmatrix}
\gamma_{4}f		& \gamma_{3}f^{*}	\\
\gamma_{1}		& \gamma_{4}f
\end{pmatrix}
H_{02} \!=\! 
\begin{pmatrix}
0				& \frac{1}{2} \gamma_2	\\
0				& 0
\end{pmatrix},\nonumber 
\end{eqnarray}
where $\gamma_0f^*\!=\!v_0\pi$ near the K point. 
Here $v_0\!=\!\sqrt{3}\gamma_0a/2\hbar$ and $\pi\!=\!\hbar(k_x\!+\!ik_y)$.  
The fitted parameter values are listed in Table~\ref{tableEM}.
Figure~\ref{figEM1} compares the original bands obtained from our DFT calculations 
and the bands obtained from the DFT-informed effective model.

\begin{table}[h]
\centering
\caption{Summary of hopping parameters in units of meV obtained by fitting DFT bands of 
tetralayer rhombohedral graphene to a low-energy effective model.}
\vspace{0.05in}
\begin{tabular}{|c|c|c|c|c|c|}
\hline
$\gamma_{0}$	& $\gamma_{1}$	& $\gamma_{2}$	& $\gamma_{3}$	& $\gamma_{4}$	& $\delta$			\\
\hline
$\;\;$3160$\;\;$	& $\;\;$445$\;\;$	& $\;\;$-18.2$\;\;$	& $\;\;$-319$\;\;$	& $\;\;$-79$\;\;$		& $\;\;$-0.066$\;\;$	\\
\hline
\end{tabular}
\label{tableEM}
\end{table}

\indent\textcolor{blue}{\em Lattice regularization.}---%
To compute the Chern number, we lattice-regularize the eight-band model with the mean-field QM potential $H_{\rm{int}}$ 
introduced in Eq.~(\ref{eq:effective_qm_int}) and the most likely mean-field pair potentials. 
For $H_{\rm{int}}$ the $\tau_z$ term can be replaced by the following Haldane-like next-nearest-neighbor hopping terms:
\begin{eqnarray}
\tau_{z} = \frac{2}{3\sqrt{3}} \sum_{n=1}^{3} \sin\left(\bm{\bar k}\cdot \bm{\beta}_{n} \right),
\end{eqnarray}
where $\bm{\beta}_{1} \!=\! -\bm{a}_{1}$, $\bm{\beta}_{2} \!=\!  \bm{a}_{1} \!-\! \bm{a}_{2}$, and $\bm{\beta}_{3} \!=\!  \bm{a}_{2}$. 
$\bm{a}_{1}$ and $\bm{a}_{2}$ are the primitive lattice vectors of graphene. 
$\bm{\bar k}$ is the crystal momentum measured from the $\Gamma$ point.
For $k_{\pm}\sigma_{\pm}$ type pair potentials, we implement the next-nearest-neighbor pairing for sublattice 1A or 4B: 
$k_{+}\sigma_{\pm} \!\rightarrow\! g(\bm{\bar k}) \sigma_{\pm}$, where
\begin{eqnarray}
g(\bm{\bar k}) = \frac{4}{3} \sum_{n =1}^{3} \omega^{n-1} \sin\left(\bm{\bar k}\cdot\bm{\beta}_{n}\right).
\end{eqnarray}

\begin{figure}[t]
\centering
\includegraphics[width=1.0\linewidth]{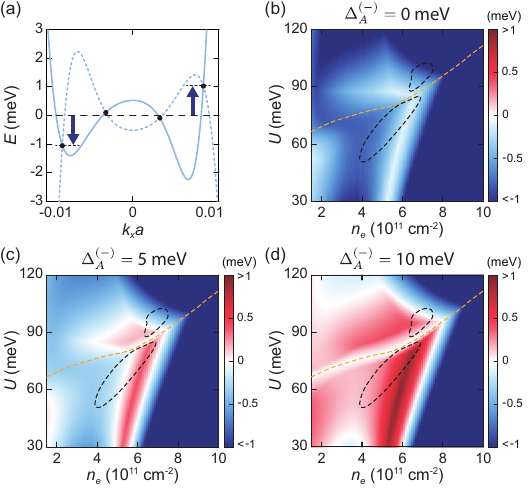}
\caption
{(a) BdG spectrum for the QM state with $U\!=\!110$~meV and $n_{e}\!=\!8\times10^{11}$~cm$^{-2}$ in Fig.~\ref{fig1}(b). 
The particle (hole) band is plotted by the solid (dashed) line. 
The arrows indicate half of the NIG, {\it i.e.}, the minimal energy required to open a global gap in the BdG spectrum.
(b) NIG map for Fig.~\ref{fig1}(b) in the absence of superconductivity.  
(c) Indirect superconducting gap map for Fig.~\ref{fig1}(b) with $\Delta_{A}^{(-)}\!=\!5$~meV. 
(d) The same as (c) but with with $\Delta_{A}^{(-)}\!=\!10$~meV.}
\label{figEM2}
\end{figure}

\indent\textcolor{blue}{\em NIG and BdG spectrum.}---%
Following the discussion in the main text, here we show the definition of NIG in Fig.~\ref{figEM2}(a) 
and three BdG spectra with $\Delta_{A}^{(-)} \!=\! 0,~5,~10$~meV in Fig.~\ref{figEM2}(b-d). 
Analyzing Fig.~\ref{figEM2} suggests that the experimentally observed QM superconducting regions I and II 
lie in the regions where the Fermi surface is simple, less trigonally warped, and only with negligible (for region I) or small (for region II) NIG.

\begin{figure}[h]
\centering
\includegraphics[width=1\linewidth]{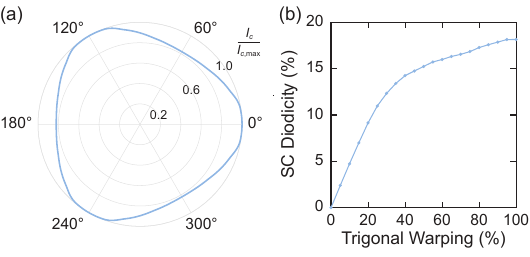}
\caption{(a) Calculated critical current $I_{c}$ versus the bias angle $\theta$.
$\theta=0$ ($\pi$) refers to the KM (K$\Gamma$) direction. 
$I_{c}$ is normalized by its maximal value $I_{c,{\rm max}}$.
(b) Calculated supercurrent diodicity versus the trigonal warping strength, 
which is set by $\gamma_2$ and $\gamma_3$ in the effective model.
The superconducting state I in Fig.~\ref{fig1}(b) with $\Delta_{A}^{(-)}\!=\!10$~meV is used here.}
\label{figEM3}
\end{figure}

\indent\textcolor{blue}{\em SDE calculation.}---%
The supercurrent diodicity $\eta(\theta)$ is defined in the main text,
and here we elaborate how we calculate the critical sheet current density $j_c$ along the bias direction $\hat{n}$, 
where $\cos(\theta)\!=\!\hat{n}\cdot\hat{n}_0$ and $\hat{n}_0$ is the KM direction.
$j_c(\hat{n})$ can be obtained~\cite{Daido2022a} from the free energy $\Omega$, a function of the Cooper-pair momentum $q$ along $\hat{n}$:
\begin{eqnarray}
j_{c}(\hat{n}) &=& \max\left\{\frac{2e}{\hbar} \partial_{q} \Omega(\Delta,q\hat{n})\right\}_{q > 0} \nonumber\\
&=& \max\left\{\frac{2e}{\hbar} \partial_{q} \frac{k_{\rm B}T}{A} \ln {\rm Tr~} e^{-\frac{\hat{H}_{\rm BdG}(\Delta,q\hat{n})}{k_{\rm B}T}}\right\}_{q > 0},
\end{eqnarray}
where $A$ is the total area of the system.
Figure~\ref{figEM3}(a) shows the anisotropy of $j_{c}$ when $\Delta_{A}^{(-)} = 10$~meV.
Markedly, in our current model calculation, by decreasing $\Delta_{A}^{(-)}$ from 10 to 5 (1)~meV, 
the maximal supercurrent diodicity increases from 18.1\% to 41.3\% (34.4\%), 
although $j_{c}$ decreases significantly. We leave this for future study. 
Figure~\ref{figEM3}(b) shows the supercurrent diodicity as a function of the strength of the trigonal warping. 
100\% refers to the parameter values of $\gamma_{2}$ and $\gamma_{3}$ listed in Table~\ref{tableEM}, 
a result of our first-principles DFT calculations. Clearly, the diodicity originates from the trigonal warping, 
which breaks the artificial continuous rotational symmetry present in the model without trigonal warping. 
Our predicted PDW diode effect may be used to determine or tested by the crystallography of graphene.

\indent\textcolor{blue}{\em Superfluid stiffness.}---% 
We calculate the superfluid stiffness from the BdG Hamiltonian using the formula
\begin{widetext}
\vspace{-0.2in}
\begin{align}
    \rho_s =\frac{1}{A}\sum_{i j \bm k} \frac{f(\epsilon_j) - f(\epsilon_i)}{\epsilon_i - \epsilon_j}
    (|\langle \psi_i|\partial_{k_x} H_{\rm BdG}|\psi_j \rangle |^2
    -|\langle \psi_i|\partial_{k_x} H_{\rm BdG}\gamma_z|\psi_j \rangle |^2),
\end{align}
\vspace{-0.2in}
\end{widetext}
which can be derived in the linear response theory~\cite{Rossi2021a,Torma2022a}.
Here $f$ is the Fermi distribution, $\bm{\gamma}$ are Pauli matrices acting on the Nambu particle-hole space,  
$\epsilon_i$ and $|\psi_i\rangle$ are the eigenvalues and eigenstates of the BdG Hamiltonian that are $\bm k$ dependent. 
Our calculations suggest that for small $\Delta$ the stiffness is mainly determined by 
the energy shell $\pm |\Delta k_F a|$ around the Fermi energy. 
Within this energy range, the Berry flux is small because of the large band gap induced by the displacement field. 
Besides the eight-band calculations that contain the full wavefunction information, 
we also perform a one-band calculation. The one-band stiffness has no geometric contribution, 
because only the lowest conduction band dispersion $E_{\bm k}$ is used (the wavefunctions are chosen to be fully sublattice polarized). 
The single-band BdG Hamiltonian reads:
\begin{align}
    H_{\rm BdG}^{(1)}=E_{\bm k}\gamma_+ - E_{-\bm k}\gamma_- - \mu \gamma_z + \Delta a(k_x \gamma_x - k_y \gamma_y).
\end{align}
For the case without the trigonal warping, we obtain the following analytical result:
\begin{equation}
    \!\!\rho_s \!=\!\frac{2}{A}\!\sum_{\bm k} \frac{\sin^2{\theta_k}(\partial_{k_x} E_k)^2
    \!-\!\Delta a\sin{2\theta_k} \cos{\phi_k} \partial_{k_x} E_k}{\epsilon_1-\epsilon_2}, 
\end{equation}
where $\sin{\theta_k} \!=\! 2\Delta ka /(\epsilon_1\!-\!\epsilon_2)$ and $\cos{\theta_k} \!=\! 2E_{k} /(\epsilon_1\!-\!\epsilon_2)$. 
The first term is the conventional intra-band contribution, 
which has the same form as the diamagnetic contribution of an ordinary superconductor. 
For a parabolic band with $s$-wave pairing, this term yields $\rho_s \!\propto\! n_s/m^*$, 
where $n_s$ is the superfluid density and $m^*$ is the effective mass. 
The second term has a momentum dependence originating from the order parameter winding, 
and it generally reduces the stiffness. 

\end{appendix}

\end{document}